\documentclass[12pt]{article}
\usepackage{amssymb,epsfig}
\newcommand{\es}{$\mathrm{E_6}$}

\begin{document}

\hspace*{\fill}UWThPh-2002-32

\hspace*{\fill}WUE-ITP-2002-028

\hspace*{\fill}hep-ph/0210363

\vspace{1cm}

\begin{center}
\textbf{\Large Singlino-dominated Neutralinos\\
in Extended Supersymmetric Models%
\footnote{Contribution to the Proceedings
of the \emph{10th International Conference on
Supersymmetry and Unification of Fundamental Interactions} (SUSY 02),
DESY, Hamburg, Germany, 17 -- 23 June 2002.}
}

\vspace{10mm}
S.~Hesselbach\footnote{e-mail: stefan.hesselbach@univie.ac.at} 

Institut f\"ur Theoretische Physik, Universit\"at Wien,
A-1090 Wien, Austria

\vspace{5mm}
F.~Franke\footnote{e-mail: fabian@physik.uni-wuerzburg.de}

Institut f\"ur Theoretische Physik und Astrophysik, 
Universit\"at W\"urzburg,\\
D-97074 W\"urzburg, Germany
\end{center}

\vspace{5mm}
\begin{abstract}
We adopt the SUSY benchmark scenario SPS 1a in supersymmetric models
where the Higgs sector is extended by singlet superfields. We consider
light neutralinos with dominant singlino character whose 
couplings are generally suppressed.
The cross sections for direct production of the singlino-dominated
neutralinos can be of the order of several f\/b for singlet vevs of some TeV.
Hence even exotic neutralinos which are not the LSP and are 
omitted in the decay chains of the other supersymmetric particles
may be visible at a high luminosity linear $e^+e^-$ collider.
If, however, the LSP is singlino-dominated the decay width of the
NLSP can be very small and displaced vertices exist for large
singlet vacuum expectation values.
\end{abstract}

\section{Introduction}

In the Next-to-Minimal Supersymmetric Standard
Model (NMSSM) \cite{nmssm1} or an \es\ inspired model with one extra
neutral gauge boson $Z'$ and one additional singlet superfield
\cite{hesselb} neutralinos with a dominant singlet higgsino (singlino)
component exist for large values of the singlet vacuum expectation
value (vev) $x \gtrsim 1$~TeV. 
We study scenarios where the MSSM-like
neutralinos have similar masses and mixing character as in the
`typical mSUGRA' SPS 1a 
scenario for the MSSM \cite{SPS}.

Since the singlino component does not couple to gauge bosons,
gauginos, (scalar) leptons and (scalar) quarks, cross sections for
the production of the exotic neutralinos are generally small. We
analyze the regions of $x$
where the associated production of the singlino-dominated neutralino
yields detectable cross sections at a linear $e^+e^-$collider. 

Since also the decay of an MSSM-like neutralino into a
singlino-dominated neutralino is strongly suppressed the decay vertex
can be significantly displaced from the production vertex in the
detector. Within the SPS 1a scenario we discuss the singlet vacuum
expectation values which lead to displaced decay vertices of an
MSSM-like next-to-lightest supersymmetric particle (NLSP) decaying
into an exotic LSP. 
The production of singlino-dominated neutralinos is extensively
discussed in \cite{singprod, singlinoprod, singlinoprodpol}. More
details of displaced vertices can be found in \cite{displaced}.

\section{Scenarios}

\subsection{NMSSM}

The NMSSM parameters (for details see \cite{nmssm1}) $M_1 = 99$~GeV,
$M_2 = 193$~GeV, $\tan\beta = 10$ and
the effective $\mu$ parameter $\mu_\mathrm{eff} = \lambda x = 352$~GeV
are chosen according to the scenario SPS 1a.
For large $x \gg |M_2|$ a singlino-dominated neutralino 
$\tilde{\chi}^0_S$ with mass
$\approx 2 \kappa x$ in zeroth approximation decouples in 
the neutralino mixing matrix
while the other four neutralinos $\tilde{\chi}^0_{1,\ldots,4}$
have MSSM character as in SPS 1a with
masses 96~GeV, 177~GeV, 359~GeV and 378~GeV.

\begin{figure}[ht!]
\begin{picture}(16,12)

\put(0,0){\epsfig{file=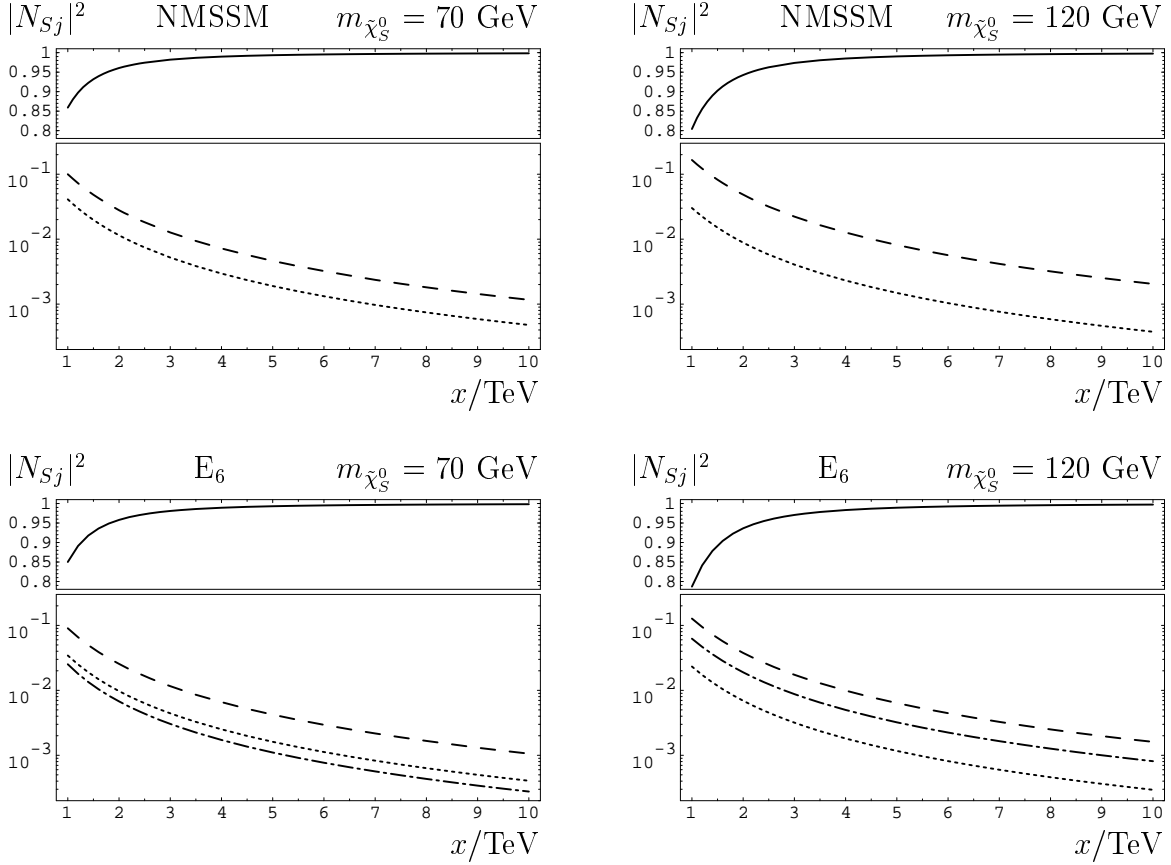,scale=1}}

\end{picture}
\caption{\label{sing} Mixing of the singlino-dominated
neutralino $\tilde{\chi}^0_S$
in the SPS 1a inspired scenarios in the NMSSM and \es{} model
with $M_1 = 99$~GeV, $M_2 = 193$~GeV, $\tan\beta = 10$ and
$\mu_\mathrm{eff} = \lambda x = 352$~GeV:
singlino content $|N_{S5}|^2$ (solid line),
MSSM gaugino content $|N_{S1}|^2 + |N_{S2}|^2$ (dashed),
MSSM doublet higgsino content $|N_{S3}|^2 + |N_{S4}|^2$ (dotted)
and $\tilde{Z}'$ content $|N_{S6}|^2$ (dashed-dotted).
The mass of $\tilde{\chi}^0_S$ is fixed at 70~GeV and 120~GeV by
the parameters $\kappa$ (NMSSM) and $M'$ (\es\ model).}
\end{figure}

Fig.~\ref{sing} shows the mixing character of $\tilde{\chi}^0_S$
as a function of the singlet vacuum expectation value $x$
while $m_{\tilde{\chi}^0_S} = 70$~GeV and 120~GeV and
$\mu_\mathrm{eff}$ are fixed
by the parameters $\kappa$ and $\lambda$, respectively. 
In order to obtain a singlino content of 90\% (99\%), $x$ must be
larger than about 1200 GeV (4000 GeV) for
$m_{\tilde{\chi}^0_S}=70$~GeV and 1500 GeV (5000 GeV) for
$m_{\tilde{\chi}^0_S}=120$~GeV.
Generally the singlino content increases with increasing value of
$x$.
The couplings of the singlino-dominated neutralino and therefore 
the production cross sections and decay widths are determined by the
remaining MSSM content, which decreases
as $1/x^2$ in good approximation.
In the SPS 1a inspired scenario the gaugino
content dominates the MSSM content of $\tilde{\chi}^0_S$,
which increases the couplings to the gaugino-like
$\tilde{\chi}^0_1$ and $\tilde{\chi}^0_2$ \cite{singprod}.

\subsection{ $\mathbf{E_6}$ model}

We consider an \es\ inspired model with one extra neutral gauge boson
$Z'$ and one additional singlet superfield which 
contains six neutralinos
\cite{hesselb}. 
Again the MSSM parameters and masses of the MSSM-like neutralinos
are fixed according to the scenario SPS 1a, while a nearly pure light 
singlino-like neutralino $\tilde{\chi}^0_S$
with mass $\approx 0.18 \,x^2/|M'|$ in zeroth approximation 
exists for very large values $|M'| \gg x$ \cite{decarlosespinosa}.
The sign of $M'$ is fixed by requiring relative sign $+1$ between the
mass eigenvalues of $\tilde{\chi}^0_S$ and $\tilde{\chi}^0_1$
\cite{singprod}.

In scenario SPS 1a with $m_{\tilde{\chi}^0_S}=70$~GeV
the $\tilde{Z}'$ content is very small (Fig.~\ref{sing}). Hence the
MSSM components of 
the singlino-dominated neutralino are of comparable size as in the
corresponding NMSSM scenario. A singlino purity of 90\,\% is reached
at $x = 1260$~GeV and 99\,\% at $x = 4200$~GeV.
In the scenario with $m_{\tilde{\chi}^0_S}=120$~GeV the $\tilde{Z}'$
content is larger resulting in a smaller gaugino content of
$\tilde{\chi}^0_S$ than in the NMSSM. Here $x$ must be larger than
1600~GeV and 5200~GeV to obtain a singlino content of 90\,\% and
99\,\%, respectively.

\section{Production of singlino-dominated neutralinos}

Direct experimental evidence of a fifth neutralino would be
an explicit proof for an extended SUSY model. An exotic
neutralino which is not the LSP is omitted in the decay cascades and
must be directly produced with sufficient cross section to be
detected.
The visibility of the cross sections for neutralinos with major exotic
components is also crucial to apply sum rules in order to test the
closure of the neutralino system \cite{choikalinowski}.

We discuss the associated production of the singlino-dominated
$\tilde{\chi}^0_S$ together with the lightest MSSM-like neutralino
$\tilde{\chi}^0_1$ in $e^+ e^-$ annihilation. Since in the SPS 1a
inspired scenario 
$\tilde{\chi}^0_1$ is gaugino-like the
production proceeds mainly via $t$- and $u$-channel exchange of
selectrons and the gaugino content of $\tilde{\chi}^0_S$ is crucial 
for the size of the cross section (Fig.~\ref{sing}).
Analytical formulae are given in \cite{franke} for the NMSSM and
\cite{hesselb} for the \es\ model.
The selectron masses are fixed according to the SPS 1a
scenario in both models: $m_{\tilde{e}_R} = 143$~GeV and
$m_{\tilde{e}_L} = 202$~GeV.
We will assume a cross section of 1~f\/b to be sufficient of the 
production process. Of course the discovery limit depends on
the neutralino decay properties that are discussed in detail
in \cite{displaced, franke, hugonie}.

\begin{figure}[bt]
\begin{picture}(16,12)

\put(0,0){\epsfig{file=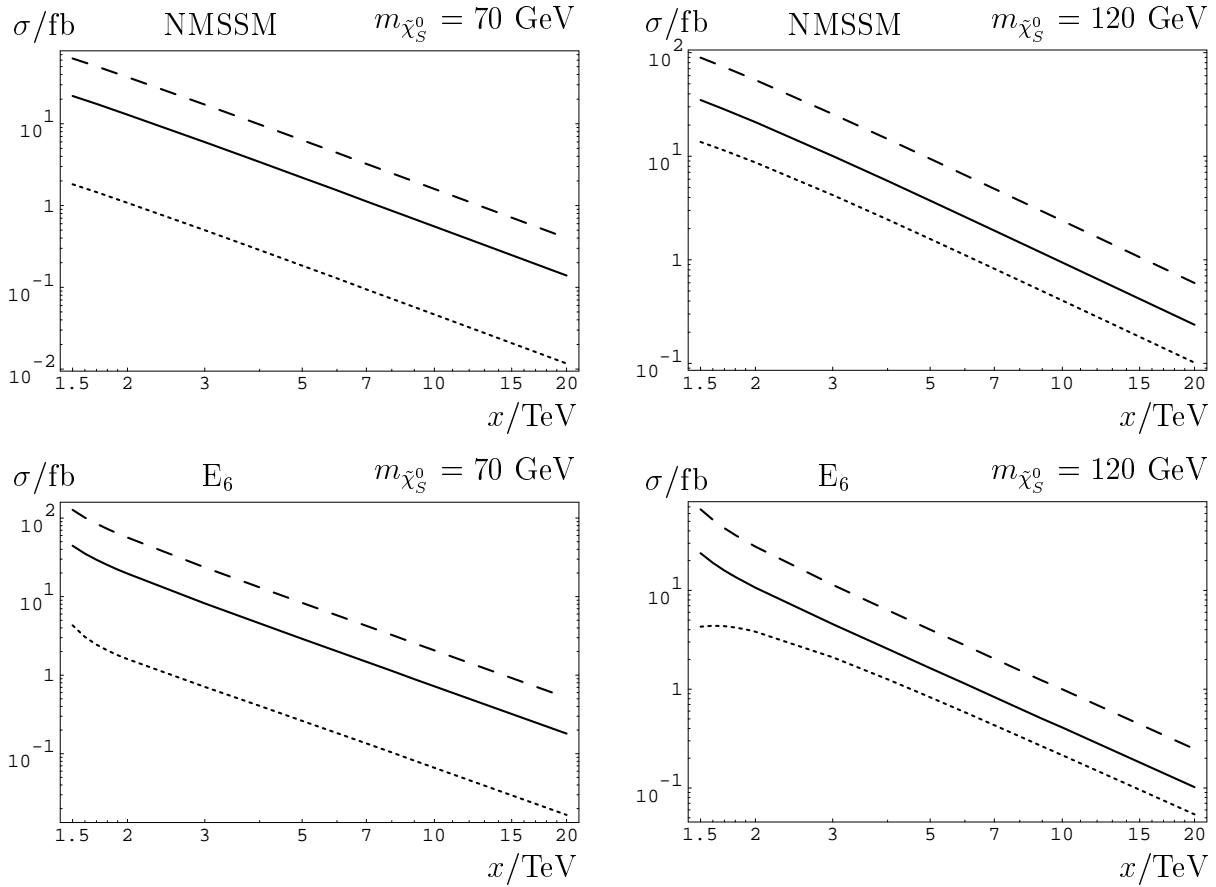,scale=1}}

\end{picture}
\caption{\label{prodplots}Cross sections for the production of a 
singlino-dominated neutralino $\tilde{\chi}^0_S$ via $e^+ e^- \to
\tilde{\chi}^0_S \tilde{\chi}^0_1$ for $\sqrt{s} = 500$~GeV
in the SPS 1a inspired scenarios in the
NMSSM and \es{} model with $M_1 = 99$~GeV, $M_2 = 193$~GeV, $\tan\beta
= 10$ and $\mu_\mathrm{eff} = \lambda x = 352$~GeV
with unpolarized beams (solid) and beam polarizations
$P_-=+0.8$, $P_+=-0.6$ (dashed) and
$P_-=-0.8$, $P_+=+0.6$ (dotted).
The mass of $\tilde{\chi}^0_S$ is fixed at 70~GeV and 120~GeV by
the parameters $\kappa$ (NMSSM) and $M'$ (\es\ model).}
\end{figure}

In Fig.~\ref{prodplots} the cross sections are shown for two masses 70
and 120~GeV of $\tilde{\chi}^0_S$, where the singlino-dominated
neutralino is the LSP and NLSP, respectively.
In all scenarios the cross sections are decreasing in good
approximation as $1/x^2$ according to the gaugino content of
$\tilde{\chi}^0_S$. 
In the NMSSM with $m_{\tilde{\chi}^0_S} = 70$~GeV (120 GeV)
the unpolarized cross
section is larger than 1~f\/b for $x< 7.4$~TeV (9.7 TeV) which
corresponds to a singlino content of 99.7\,\%. 
While the cross sections in the \es{} model 
with $m_{\tilde{\chi}^0_S} = 70$~GeV 
are enhanced because of a positive $\tilde{Z}'$ contribution
to the electron-selectron-neutralino coupling,
the smaller gaugino content of $\tilde{\chi}^0_S$
and a negative $\tilde{Z}'$ contribution
reduces the cross section for $m_{\tilde{\chi}^0_S} = 120$~GeV.

We also show the cross sections for polarized beams with 80\,\%
electron and 60\,\% positron polarizations expected at a future linear
collider \cite{pol}. In the extended SPS 1a scenarios always the
configuration $P_-=+0.8$, $P_+=-0.6$ enhances the cross sections by a
factor 2 -- 3 \cite{singlinoprodpol, gudi}.

\section{Displaced vertices}

Displaced vertices of an MSSM-like NLSP
appear if the couplings of a singlino-dominated LSP to the NLSP
are strongly suppressed at large values of $x$ \cite{displaced, hugonie}. 
Then assuming conserved $R$-parity all SUSY particles first decay into
the NLSP which finally decays into the LSP with a decay
vertex displaced from the production vertex.
 
\begin{figure}[bt]
\begin{picture}(16,6)

\put(0,0){\epsfig{file=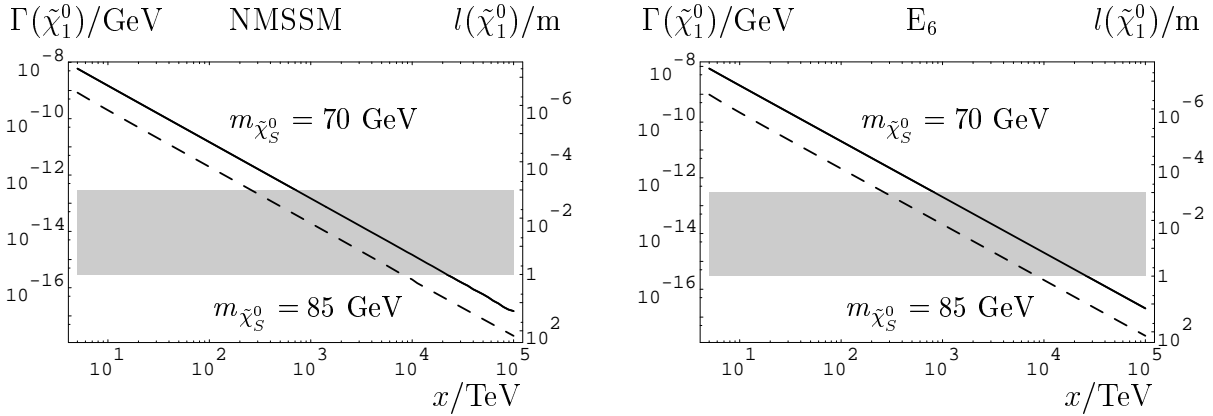,scale=1}}

\end{picture}
\caption{\label{displacedplots}Total decay widths of the lightest
MSSM-like neutralino $\tilde{\chi}^0_1$ decaying into a
singlino-dominated neutralino $\tilde{\chi}^0_S$
in the SPS 1a inspired scenarios in the 
NMSSM and \es{} model with $M_1 = 99$~GeV, $M_2 = 193$~GeV, $\tan\beta
= 10$ and $\mu_\mathrm{eff} = \lambda x = 352$~GeV.
The mass of $\tilde{\chi}^0_S$ is fixed at 70~GeV and 85~GeV by
the parameters $\kappa$ (NMSSM) and $M'$ (\es\ model).
The shaded area marks the decay widths where displaced
vertices exist. Below this area the decaying particle
escapes detection.}
\end{figure}

In Fig.~\ref{displacedplots} we show the total decay width and decay
length of the MSSM-like NLSP $\tilde{\chi}^0_1$ for two masses of the
singlino-like
LSP $\tilde{\chi}^0_S$ of 70~GeV and 85~GeV. 
For the larger mass difference between $\tilde{\chi}^0_S$ and
$\tilde{\chi}^0_1$ displaced vertices appear for $x>7\cdot
10^2$~TeV and $x>8\cdot 10^2$~TeV in the NMSSM and \es{} model,
respectively. For the smaller mass difference phase space effects
outweigh the larger mixing and the area of displaced vertices is
reached for $x>3\cdot 10^2$~TeV \cite{displaced}.

Since the direct production of singlino-like neutralinos is visible up
to $\mathcal{O}(10~\mathrm{TeV})$, displaced vertices would be helpful
to study exotic LSPs for large $x = 10^2$ -- $10^4$~TeV. If, however,
the $\tilde{\chi}^0_S$ is not the LSP it remains invisible in this
parameter region. 

\section{Conclusion}
We have discussed the production of singlino-dominated neutralinos
and the neutralino decay with displaced vertices in extended versions of the
SPS 1a scenario within the NMSSM and an \es\ inspired model with one new
gauge boson.  
Since the singlet vacuum expectation value $x$ determines the singlino
content it is the crucial parameter which triggers the
production and decay properties of the exotic neutralino.

An exotic neutralino that is not the LSP is only visible in direct 
production since it is omitted in the decay cascades of the other SUSY
particles. At a high luminosity linear $e^+e^-$ collider the
associated production of a singlino-dominated neutralino is detectable
up to $x$ values of order 10~TeV, which corresponds to a singlino
content of more
than 99\,\%. Displaced vertices of an MSSM-like NLSP into a 
singlino-dominated LSP exist for very large $x$ of order 100 --
1000~TeV, significantly above the region where direct production occurs.
 
\section*{Acknowledgment}

We thank A.~Bartl and H.~Fraas for many helpful discussions and the
careful reading of the manuscript.
This work is supported by the `Fonds zur F\"orderung der
wissenschaftlichen For\-schung' of Austria, FWF Project No.~P13139-PHY,
by the EU TMR Project No.\ HPRN-CT-2000-00149
and by the Deutsche Forschungsgemeinschaft (DFG) under contract No.\
\mbox{FR~1064/5-1}.

\end{document}